\documentclass[usenatbib, longauth]{aa} 

\usepackage{graphicx}
\usepackage{graphics}
\usepackage{txfonts}
\usepackage{epsfig}
\usepackage{natbib}
\usepackage{float}
\usepackage{caption}
\usepackage{hyperref}
\usepackage{dblfloatfix}
\usepackage[]{color}
\begin{document}

\title{The gas streamer G1-2-3 in the Galactic Center}

\titlerunning{The Galactic Center gas-streamer G-1-2-3}
\subtitle{}

\author{
   S.~Gillessen\inst{1}
   \and F.~Eisenhauer\inst{1,3}
\and  J.~Cuadra\inst{4,9}
   \and R.~Genzel\inst{1,2}
\and   D.~Calderon\inst{5}
\and S.~Joharle\inst{1}
\and T.~Piran\inst{7}
\and D.C.~Ribeiro\inst{1}
\and C.M.P.~Russell\inst{8}
\and M.~Sadun Bordoni\inst{1}
 \and A.~Burkert\inst{6,1}
 \and   G.~Bourdarot\inst{1}
\and A.~Drescher\inst{1}
\and F.~Mang\inst{1}
\and T.~Ott\inst{1}
\and\\ G.~Agapito\inst{10}
\and A.~Agudo~Berbel\inst{1}
\and A.~Baruffolo\inst{12}
\and M.~Bonaglia\inst{10}
\and M.~Black\inst{13}
\and R.~Briguglio\inst{10}
\and Y.~Cao\inst{1}
\and L.~Carbonaro\inst{10}
\and G.~Cresci\inst{10}
\and Y.~Dallilar\inst{16}
\and R. Davies\inst{1}
\and M.~Deysenroth\inst{1}
\and I.~Di~Antonio\inst{11}
\and A.~Di~Cianno\inst{11}
\and G.~Di~Rico\inst{11}
\and D.~Doelman\inst{15}
\and M.~Dolci\inst{11}
\and S.~Esposito\inst{10}
\and D.~Fantinel\inst{12}
\and D.~Ferruzzi\inst{10}
\and H.~Feuchtgruber\inst{1}
\and N.~M.~F\"orster~Schreiber\inst{1}
\and A.~M.~Glauser\inst{14}
\and P.~Grani\inst{10}
\and M.~Hartl\inst{1}
\and D.~Henry\inst{13}
\and H.~Huber\inst{1}
\and C.~Keller\inst{15}
\and M.~Kenworthy\inst{15}
\and K.~Kravchenko\inst{1}
\and J.~Lightfoot\inst{13}
\and D.~Lunney\inst{13}
\and D.~Lutz\inst{1}
\and M.~Macintosh\inst{13}
\and F.~Mannucci\inst{10}
\and D.~Pearson\inst{13}
\and A.~Puglisi\inst{10}
\and S.~Rabien\inst{1}
\and C.~Rau\inst{1}
\and A.~Riccardi\inst{10}
\and B.~Salasnich\inst{12}
\and T.~Shimizu\inst{1}
\and F.~Snik\inst{15}
\and E.~Sturm\inst{1}
\and L.~J.~Tacconi\inst{1}
\and W.~Taylor\inst{13}
\and A.~Valentini\inst{11}
\and C.~Waring\inst{13}
\and M.~Xompero\inst{10}
}

\institute{
Max Planck Institute for extraterrestrial Physics,
Giessenbachstra{\ss}e~1, 85748 Garching, Germany
\and Departments of Physics and Astronomy, Le Conte Hall, University
of California, Berkeley, CA 94720, USA
\and Technical University of Munich, 85747 Garching, Germany
\and Universidad Adolfo Iba\~nez, Av. Padre Hurtado 750, Vi\~na del Mar, Chile 
\and Max Planck Institute for Astrophysics, Karl-Schwarzschild-Stra{\ss}e~1, 85748 Garching, Germany
\and University Observatory Munich, Scheinerstra{\ss}e 1, 81679 Munich, Germany
\and Racah Institute for Physics, The Hebrew University, Jerusalem, 91904, Israel
\and Department of Physics and Astronomy, Bartol Research Institute, University of Delaware, Newark, DE 19716, USA
\and Millennium Nucleus on Transversal Research and Technology to Explore Supermassive Black Holes (TITANS), Chile
\and INAF -- Osservatorio Astrofisico di Arcetri, Largo E. Fermi 5., 50125, Firenze, Italy
\and INAF -- Osservatorio Astronomico d'Abruzzo, Via Mentore Maggini, 64100, Teramo, Italy
\and INAF -- Osservatorio Astronomico di Padova, Vicolo dell'Osservatorio 5, 35122, Padova, Italy
\and STFC UK ATC, Royal Observatory Edinburgh, Blackford Hill. Edinburgh, EH9 3HJ, UK
\and ETH Zurich, Institute of Particle Physics and Astrophysics, Wolfgang-Pauli-Strasse 27, 8093 Zurich, Switzerland
\and Leiden Observatory, University of Leiden, P.O. Box 9513, 2300 RA Leiden, The Netherlands
\and I. Physikalisches Institut, Universit\"at zu K\"oln, Z\"ulpicher Str. 77, 50937, K\"oln, Germany
}

\date{Draft version \today}

\abstract{
The black hole in the Galactic Center, Sgr~A*, is prototypical for ultra-low-fed galactic nuclei. The discovery of a hand-full of gas clumps in the realm of a few Earth masses in its immediate vicinity provides a gas reservoir sufficient to power Sgr~A*. In particular, the gas cloud G2 is of interest due to its extreme orbit, on which it passed at a pericenter distance of around $100\,$AU and notably lost kinetic energy during the fly-by due to the interaction with the black hole accretion flow. 13 years prior to G2, a resembling gas cloud called G1, passed Sgr~A* on a similar orbit. 
The origin of G2 remained a topic of discussion, with models including a central (stellar) source still proposed as alternatives to pure gaseous clouds. Here, we report the orbit of a third gas clump moving again along (almost) the same orbital trace. Since the probability of finding three stars on close orbits is very small, this strongly argues against stellar-based source models. Instead, we show that the gas streamer G1-2-3 plausibly originates from the stellar wind of the massive binary star IRS16SW. This claim is substantiated by the fact that the small differences between the three orbits -  the orientations of the orbital ellipses in their common plane as a function of time - are consistent with the orbital motion of IRS~16SW.
}


\maketitle

\section{Introduction}
\label{sec:intro}

\subsection{Observed properties of G2}

The peculiar object G2 in the Galactic Center was detected in \citet{2012Natur.481...51G} as an object apparently residing between the S-stars \citep{1996Natur.383..415E, 1998ApJ...509..678G, 2002Natur.419..694S, 2003ApJ...586L.127G, 2005ApJ...628..246E, 2008ApJ...689.1044G, 2009ApJ...692.1075G, 2017ApJ...837...30G, 2024A&A...692A.242G}. G2 showed the properties of a dusty, ionized gas cloud: Line emission from hydrogen and helium, and thermal emission corresponding to $600\,$K. No emission from a potential stellar photosphere (with $T\gtrsim 3000\,$K) was convincingly detected, as the claimed K-band detection in \citet{2013A&A...551A..18E} remained episodic. G2 was spatially marginally resolved (at the $\sim 60\,$mas resolution of the 8m VLT in K-band), but more importantly showed spectrally a velocity gradient across the source. 

Multi-year adaptive optics imaging (with NACO, \citealt{1998SPIE.3354..606L, 1998SPIE.3353..508R}) and integral field spectroscopy (with SINFONI \citealt{2003SPIE.4841.1548E, 2003SPIE.4839..329B}) in the near-infrared showed that G2 was approaching Sgr~A* on a very eccentric orbit with a pericenter passage in 2014 \citep{2014ApJ...796L...8W, 2013ApJ...774...44G}. Position-velocity diagrams of the following years showed a spectacular tidal evolution of the gas stretching along the orbital trace \citep{2015ApJ...798..111P, 2017ApJ...840...50P}.

Given the measured size of around $120\,$AU and the thermal luminosity and temperature, it was clear that G2 is optically thin and fully ionized by the UV photons of the surrounding young, massive stars.  From the Brackett-$\gamma$ luminosity, and assuming case B recombination theory, the gas mass of G2 was estimated to be $\lesssim 3$ Earth masses.

With the gas of G2 nearly radially falling in, a number of papers addressed the question, whether there would be a radiative reaction observable around pericenter passage \citep{2012Natur.481...51G,2012ApJ...757L..20N,2013MNRAS.436.1955C,2013PhRvL.110v1102B}. The absence of any such events does, however, not yield any strong constraint on  possible source models. \citet{2015MNRAS.454.1525P} report a mildly increased rate of X-ray flares from Sgr~A* post-pericenter passage of G2.

\citet{2015ApJ...798..111P} noted that 12 years ahead of G2, a very similar, although a bit fainter, object was moving along a very similar trajectory, which they called G1. The dust emission of this object was also noted already in very early adaptive-optics images \citep{2005A&A...439L...9C}, where it appeared spatially extended \citep{2005ApJ...635.1087G}, and also showed tidal evolution along the orbit \citep{2017ApJ...847...80W}. Further, a tail follows behind G2 on apparently a similar orbit.  One could thus think of G1, G2 and the tail being knots in a  long gas streamer, of which we detect in sensitivity-limited observations only the brightest peaks, aided by the fact that the observed surface brightness of the gas scales like density squared.

The long gas streamer actually is visible in deep spectroscopic integrations (see figures~9 and~10 in \citealt{2017ApJ...840...50P}). It points spatially and in radial velocity back to the stellar wind of the contact binary system IRS~16SW \citep{1999ApJ...523..248O,2006ApJ...649L.103M}. That star is classified as Ofpe/WN9 type and a member of the clockwise (CW) disk of young stars \citep{2003ApJ...590L..33L,2006ApJ...643.1011P,2009ApJ...690.1463L,2009ApJ...697.1741B,2014ApJ...783..131Y,2022ApJ...932L...6V, 2023ApJ...949...18J}, and G2 and G1 orbit Sgr~A* in that same plane. In hydrodynamical simulations \citep{2006MNRAS.366..358C, 2020ApJ...888L...2C} the star is a significant contributor to the gas found around Sgr~A*, and observationally its spectrum shows signs of strong winds.

The post-pericenter motion of G2 is actually better described by adding a ram-pressure-like drag force to the Keplerian model \citep{2015ApJ...798..111P,2016MNRAS.455.2187M,2017MNRAS.465.2310M,2019ApJ...871..126G}. G2 decelerates due the interaction with the accretion flow of Sgr~A*\citep{2003ApJ...598..301Y,2014ARA&A..52..529Y}, the density of which follows roughly a radial profile $\propto r^{-1}$. G2's motion provides thus a valuable measurement of the accretion flow density at around $10^3$ Schwarzschild radii $r_S$, which otherwise is constrained mostly around $10\,r_S$ from submm observations \citep{2000ApJ...538L.121A,2000ApJ...539..809Q,2003ApJ...588..331B,2006ApJ...640..308M,2007ApJ...654L..57M,2015ApJ...802...69B}, or around $10^5\,r_S$ from X-ray observations \citep{2003ApJ...591..891B,2013Sci...341..981W}.

\subsection{Source models for G2}

Given the strong evidence for gaseous, tidal evolution in G2, it is clear that the observed gas cannot be gravitationally bound to a central source. With the measured size of G2 ($\approx 20\,$mas) around the time of detection \citep{2012Natur.481...51G}, the mass required to bind gas on that scale would  exceed $10^4\,M_\odot$, i.e. an intermediate mass black hole, which is excluded at such small radii \citep{2009ApJ...692.1075G,2020ApJ...888L...8N,2020ApJ...892...39R,2023A&A...672A..63G}. Yet, this does not exclude the possibility of a central stellar source that could be the origin of the gas.

Observationally, there is no need to invoke a central object for G2. All its properties can be explained without. It only has an L-band counterpart, but is invisible in K-band. Also the post-pericenter presence and appearance of G2 is not a strong discriminator, as purely gaseous models do not predict complete disruption of the cloud, see for example fig.~2 in \citet{2015ApJ...811..155S}.
Still, models with a central source have been discussed and often favored \citep{2013ApJ...773L..13P,2014ApJ...796L...8W}. \citet{Scoville_2013} proposed a young low-mass star, the wind of which would create G2. \citet{2012A&A...546L...2M} discussed a stellar nova as origin for G2. Also,  tidal interactions of a star with Sgr~A* have been proposed, either from a disk around the star \citep{2012NatCo...3.1049M,2012ApJ...756...86M} or a true, partial tidal disruption of a giant star \citep{2014ApJ...786L..12G}.

Perhaps most extreme seems the proposal, that not only G2, but maybe also other, similar gas knots \citep{2020Natur.577..337C} are binary merger products \citep{2014ApJ...796L...8W}, driven by the Kozai-Lidov mechanism \citep{2015ApJ...799..118P, 2016MNRAS.460.3494S}. This picture requires a number of assumptions, from the presence and merging of binaries close to Sgr~A*, to the question of how such an object would appear, and for how long it would do so, and how likely it is to observe a number of them simultaneously, given that the total number of stars from which the mergers can come is only larger by perhaps a factor of 20.

Here, we present new observational evidence supporting the gas streamer picture: The emission that formed the outer tail at the time of discovery of G2 has now shaped up to a clump that resembles very much G2 in 2008 - constituting thus a third object, which is moving along almost the same trajectory.
Naturally, this clump could be called 'G3', however this name has already been taken by  \citet{2020Natur.577..337C} for another gaseous object. Given the properties of G1, G2 and the new G3, it is anyhow more natural to think of them together as a gas streamer, which we will call G1-2-3. For the purpose of this paper G3 refers to the third gas clump in this series, but we propose to refer to it in future work as 'third clump of G1-2-3'.

\section{Observations}
\label{sec:observations}

\begin{figure}[h]
\centering
\includegraphics[width=9cm]{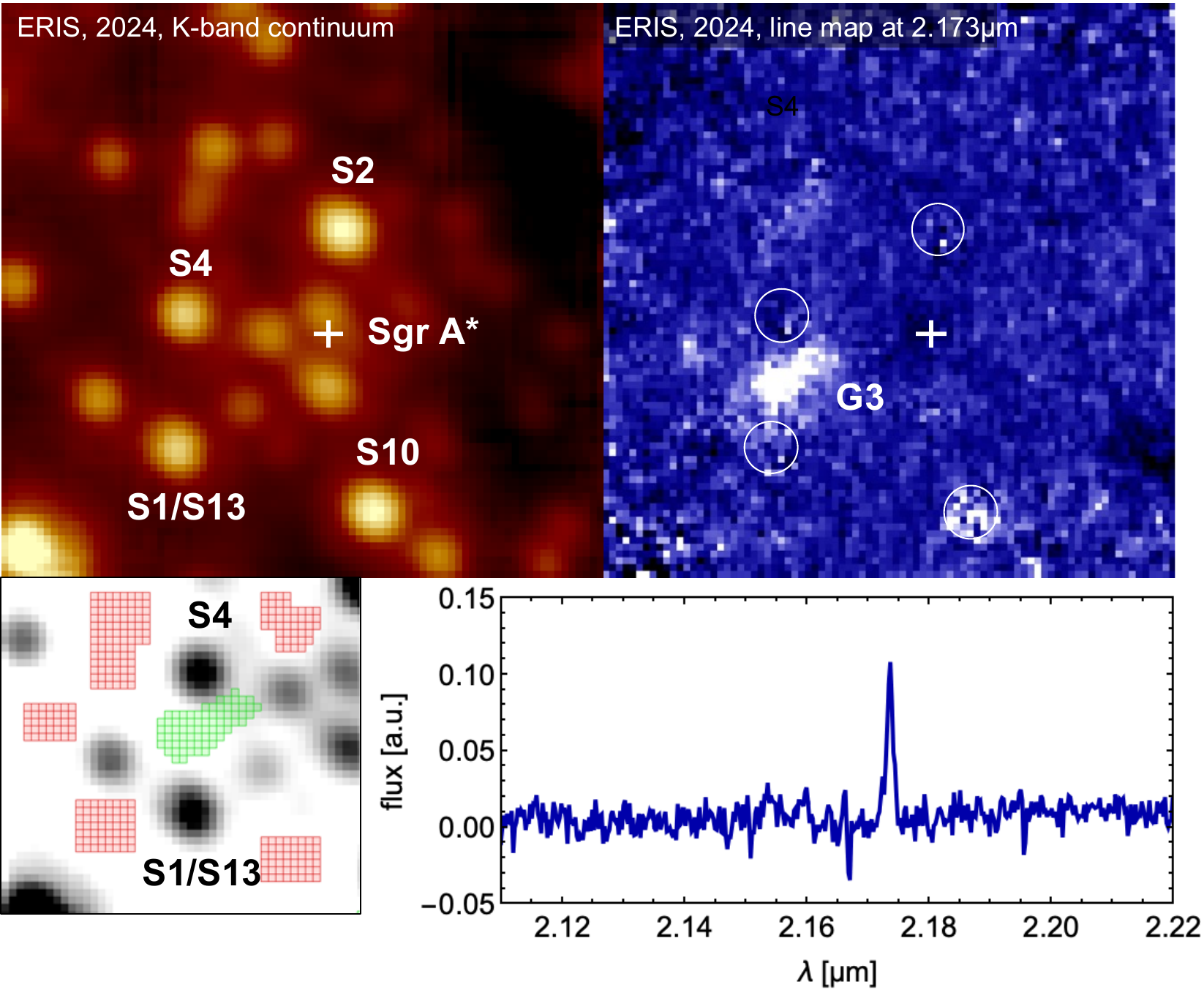}
\caption{G3 in the ERIS integral-field data from summer 2024. Top left: Continuum image showing the S-stars. Top right: Background-subtracted line map centered at $2.173\,\mu$m, corresponding to Brackett-$\gamma\, + 1000\,$km/s. G3 stands out. Bottom left: Example of a pixel selection (on - green, off - red) for extracting the G3 spectrum overlaid on the continuum map. Bottom right: The resulting spectrum shows a strong emission line at $2.173\,\mu$m.}
\label{fig1}
\end{figure}

Our data are based on adaptive-optics-assisted integral-field spectroscopy, pre-2020 with SINFONI, and since 2022 with ERIS \citep{2023A&A...674A.207D}. Both instruments use the same integral-field unit (IFU). We employ the $25\,$mas pixel scale and observe the K-band, in which we will concentrate in the analysis on the emission of Brackett-$\gamma$ (vacuum rest wavelength $2.16612\,\mu$m).  Since these data also serve the purpose of monitoring radial velocities of the S-stars, we cover a field of view of around $\pm0.6$'' by dithering the $0.8$'' field-of-view IFU into four quadrants located $\pm0.2$'' offset from Sgr~A* in both coordinates. G3 is located around $(\Delta \alpha, \Delta \delta) = (+0.3,-0.1)''$ from Sgr~A*, comfortably in the data cubes. The top row in figure~\ref{fig1} illustrates the ERIS data cube from summer 2024.

The exposure time is $600\,$s per individual data cube. Given the faintness of the line emission from G1-2-3, we typically add the data from one observing campaign lasting a few nights into a combined cube in order to reach the fainter magnitudes. Prior to adding, we align the cubes in all three dimensions, neglecting the small change of the local standard of rest (LSR). Table~\ref{tab:datalist} in the appendix summarizes the data used for G3.

\section{Analysis}
\label{sec:analysis}

\subsection{The orbits of G1, G2 and G3}

Using QFitsView \citep{qfv}, we identify the G3 emission around Brackett-$\gamma$. For obtaining an estimate of G3's orbit, we measure the position and 
radial velocity of the emission line. Since it is extended both spatially and spectrally, this is done manually, trying to capture the major part of the emission around its peak. The bottom row of figure~\ref{fig1} shows an example for the data cube from summer 2024. The spectral and spatial line positions are obtained via simple Gaussian fits. Using the pixel scale of $12.5\,$mas/pix, positions are referenced to Sgr~A* via S2, the orbit of which is known to much better precision than what is needed for G3. For the radial velocities we apply the standard LSR correction. Due to the extended nature we assign conservative error bars of $10\,$mas spatially (at least) and $50\,$km/s spectrally.  We obtain positions from six, and radial velocities from eight epochs between 2014 and 2025 in this way. For G2 we obtain compared to \citet{,2019ApJ...871..126G} in the same way five more ERIS-based radial velocities and four more positions, yielding a total of 23 velocities and 18 positions.

\begin{figure}[h]
\centering
\includegraphics[width=9cm]{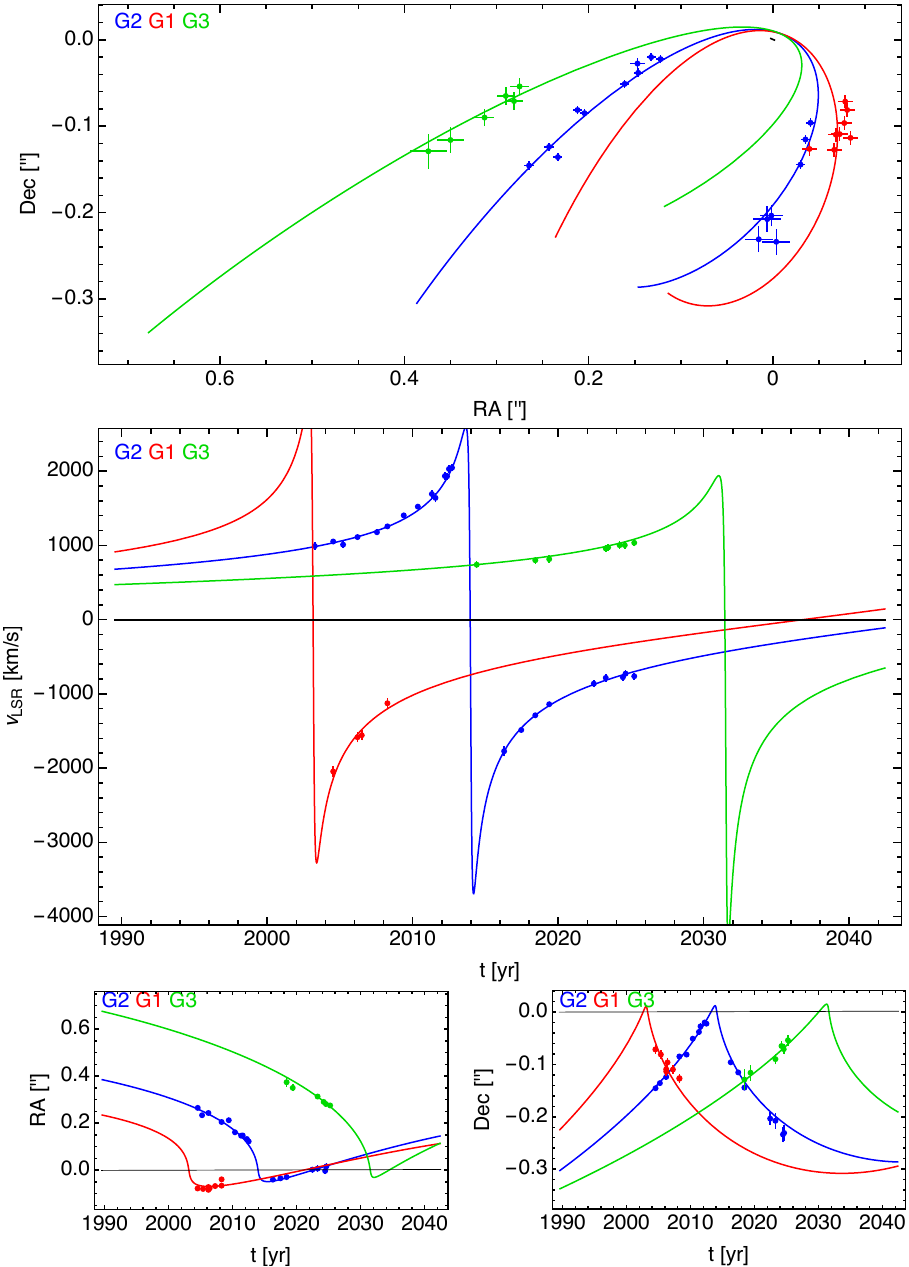}
\caption{The orbits and data of G1, G2 and G3, fit under the side constraint that they share the same orbital plane, semi-major axis and eccentricity. Top: On-sky appearance of the two orbits. Middle: The radial velocities.  Bottom: The spatial coordinates as a function of time. The non-Keplerian shapes are a result of the drag force acting on G2 and G3 when they are close to Sgr~A*.}
\label{fig3}
\end{figure}

At this point, the G3 astrometry shows no significant acceleration towards Sgr~A*, but a $>9\sigma$ significant change of radial velocity is observed, such that we have enough dynamic quantities to determine an orbit (with six degrees of freedom) in a fixed gravitational potential. The data are shown in figure~\ref{fig3}. A preliminary orbit fit for G3 (table~\ref{tab:oelist1}) shows that its orbital plane, orientation, shape and size are similar to the one of G2 (and thus also close to the CW disk and G1, \citealt{2015ApJ...798..111P}). Finding three objects by chance on such similar orbits is very unlikely: The probability for two orbital planes to agree to within $\pm\theta$ is $p_1=2\pi(1-\cos\theta)/4\pi$; the probability that the orientations of the ellipses in the plane agree to within  $\pm\theta$ is $p_2=\theta/\pi$. Finding three orbital planes has thus a  probability of $p=p_1^2 p_2^2$, which evaluates to $p=2 \times 10^{-6}$ for $\theta=15^\circ$. Similarity in orbital phase, eccentricity, and semi-major axis further reduce $p$. Hence, one can exclude a random alignment.

This  is the  motivation for our proposed  model:  The three clumps share (almost) identical orbits, due to their common origin from IRS~16SW. We thus fit the three orbits with strong side constraints: They need to be coplanar, and they need to share a common semi-major axis and eccentricity. The only differences allowed are thus the time of pericenter passage, and the longitude of periastron corresponding to the orientation of the orbital ellipses in their common plane. The gravitational potential is taken from \citet{2022A&A...657L..12G} and kept fixed ($M =4.296 \times 10^6\, $M$_\odot,\, R_0 = 8.275\,$kpc). The drag force model is used (and fit for) as in \citet{2019ApJ...871..126G}.
The resulting model is shown on top of the data in figure~\ref{fig3}. It is a very good description of the data, which is a very remarkable finding. Fitting the G2, G1, and G3 individually with each fit having six free parameters (and G2 a drag force parameter in addition) leads to $\chi^2$ values of 115.6, 36.5, and 26.4 for 50, 14, and 14 degrees of freedom respectively. This is a total $\chi^2$ of 178.5. The combined fit is only slightly worse with $\chi^2 = 196.6$, despite having eight free parameters less.
For another representation of the combined fit see figure~\ref{fig10}, and section~\ref{sec:C} for the orbital elements. 

The apocenter distance of the orbits is around 1.9'', occurring around the mid of the $19^\mathrm{th}$ century. The apocenters are not far away from where IRS~16SW was at the time (using the orbit from \citealt{2017ApJ...837...30G}). An exact agreement is anyhow not expected, as the gas clumps originate from the winds of IRS~16SW (or shocks created by them), which themselves have a significant velocity of a few hundred km/s. Also the orbital planes do not agree perfectly, but differ by $\lesssim 30^\circ$, which again might just reflect the fact that the initial conditions of the gas clouds do not need to be exactly the ones of the star. If fit individually, G2 would attain a larger apocenter distance, and G3 a smaller one. This indeed might be more realistic, as individual clumps arising in the stellar wind are free to have different initial conditions. In that view, also the orbital planes can differ a bit between the three clumps and the star. Also. previous work found that the best fits of the orbits of G1 and G2 do not completely agree \citep{2015ApJ...798..111P, 2017ApJ...847...80W}. But
 the point of the combined fit employed here is that it is remarkable, that the model in which G1 and G3 only have two free parameters each (pericenter time and longitude of pericenter) describes the data that well.

Using the combined fit, G3's pericenter passage will occur $17.6\pm0.3\,$yr after the one of G2, i.e. in mid 2031. The longitudes of pericenter differ by $12.9\pm 1.1^\circ$ between G2 and G3, with the G3 orbit being rotated clockwise against the G2 orbit. These two numbers correspond to an angular speed of $0.74 \pm 0.07 ^\circ/$yr.
For G1 in comparison with G2, the same comparison yields an angular speed of $0.74 \pm 0.10 ^\circ/$yr in the same direction (consistent with the value implied in  \citet{2015ApJ...798..111P} of $0.95\pm 0.63 ^\circ/$yr). Note that the numbers happen to be very close to each other, this was not a constraint for the fit. Further, the value is remarkably close to the angular speed of IRS~16SW, which is around $1.11 \pm 0.32^\circ$/yr at its orbits around 1950. Thus, {\bf the differences between the G1, G2 and G3 orbits can be explained by the orbital motion of IRS~16SW}. This is an additional hint towards the star being the origin of the gas clouds. We note that other (clockwise disk) stars in the vicinity could in principle also be responsible for the production of the gas clumps. Here we limit ourselves to discussing IRS~16SW, which we consider to be the most likely candidate.

\subsection{Position velocity diagram for G3}

\begin{figure}[h]
\centering
\includegraphics[width=4.0cm]{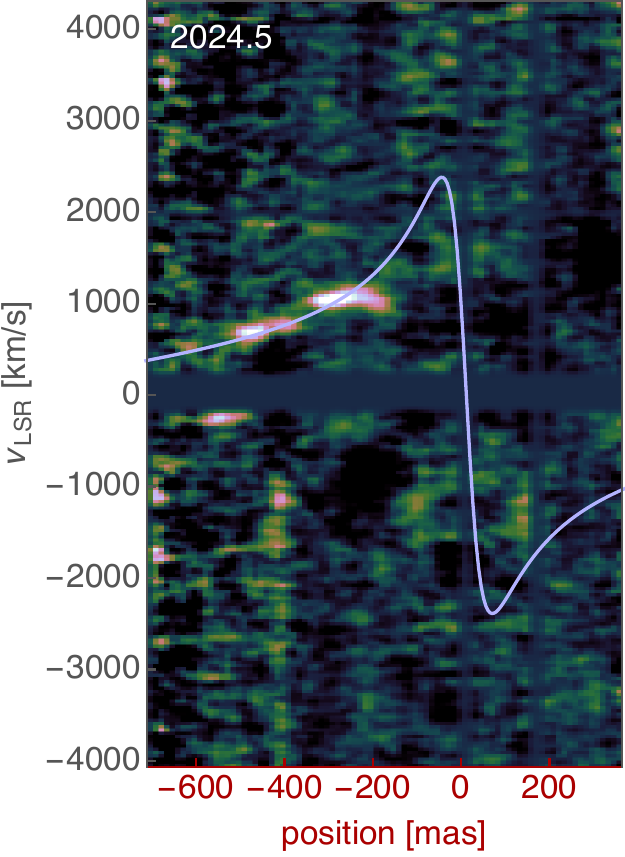}
\includegraphics[width=4.0cm]{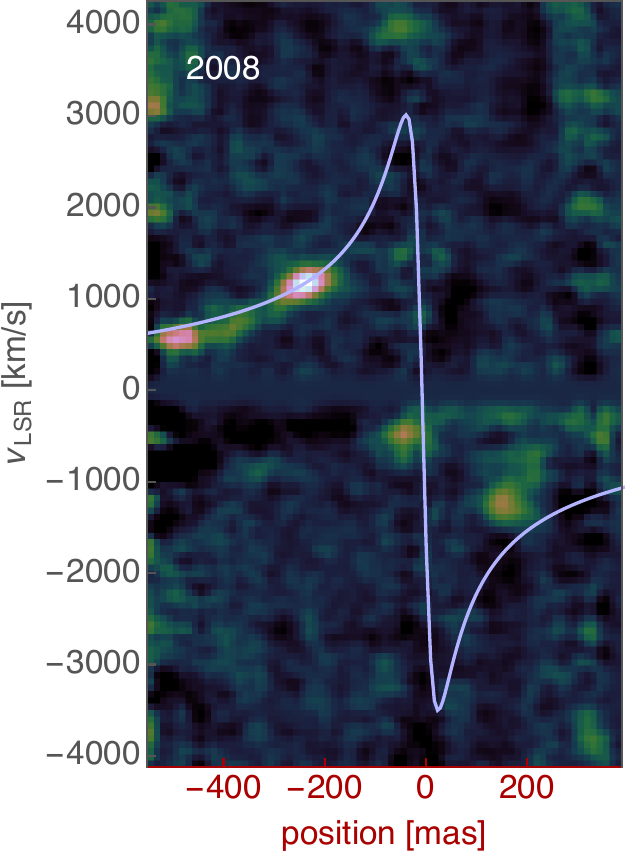}
\caption{Left: Position-velocity diagram extracted from the summer-2024 data cube, using a curved slit along the orbital trace of G3. The emission of G3 is concentrated around $(-300\,\mathrm{mas},+1000\mathrm{km/s})$. Like G2, G3 seems to be followed by a tail, indicative of even more material flowing along the G1-2-3 path. Right: For comparison, the same diagram for G2 extracted from the 2008 data cube (from \citealt{2019ApJ...871..126G}). }
\label{fig4}
\end{figure}

Using the same technique as in \citet{2019ApJ...871..126G} of extracting a position-velocity diagram along the curved trajectory given by the orbit, we obtain the diagram shown in figure~\ref{fig4}. The G3 emission resembles very much G2 a few years before pericenter passage. Notably, also G3 shows trailing emission, like G2 showed a tail as noted already in the discovery paper~\citet{2012Natur.481...51G}. The Brackett-$\gamma$ luminosity of G3 is comparable to that of G2, although roughly 38\% lower.

\section{Discussion}
    With the ample hints of G1-2-3 being related to IRS~16SW, it is worth re-considering the formation of gas clumps from the winds of that source. 
    This requires the production of clouds with masses of $\lesssim$ few Earth masses at intervals of 10--20 years, yielding a time-averaged mass rate on the order of ${\sim}10^{-7}~\text{M}_\odot~\text{yr}^{-1}$. 
    This constitutes a non-negligible fraction of around 1\% of the Wolf-Rayet wind rate of ${\sim}10^{-4.7}~\text{M}_\odot~\text{yr}^{-1}$ \citep{2007A&A...468..233M}. 
    Channelling such a substantial fraction of an initially quasi-spherical outflow toward the central SMBH poses a significant dynamical challenge as the gas  has to lose most of its orbital angular momentum. 
    An anisotropic mass-loading geometry, possibly governed by the orientation of IRS~16SW  binary axis relative to its orbital motion, may be required to facilitate such an efficient mass transfer. 
    Unfortunately, the orientation of the orbit is unknown, except that the system is close to  edge-on since it is observed as an eclipsing binary \citep{2006ApJ...649L.103M}. 

    An alternative is to consider the interaction between stellar winds and the medium. 
    In the central light-month around Sgr~A*, we expect the formation of clumps in the shocks produced by stellar winds, where a dense slab can form. 
    This slab, depending on its radiative properties, can be prone to cooling and hydrodynamic instabilities that result into the formation of clumps \citep{1994ApJ...428..186V}. 
    This process has been studied analytically and numerically for stellar wind collisions in binaries and random stellar encounters \citep{2016MNRAS.455.4388C,2020MNRAS.493..447C} . 
    These works found that for the Galactic Center stars, the clumps formed are smaller and lighter than those observed. 
    Additionally, \cite{2018MNRAS.478.3494C} investigated the hypothesis of binaries, specifically focusing on IRS~16SW, launching clumps into the medium and following their ballistic orbits. 
    The results showed that it was not possible to reproduce the position and velocity of G2 provided it formed from the binary. 
    Interestingly, the simulations of  \citet{2020ApJ...888L...2C,2025A&A...693A.180C} of the whole Wolf-Rayet population in the Galactic centre revealed a different formation mechanism that actually relied on the hydrodynamic interaction with the medium, namely, the disruption of the wind bow shock. 
    For that, the medium needs to be relatively dense, and the stellar wind needs to have a low terminal velocity, typically 450~km~s$^{-1}$. 
    
    Previous models of the Galactic centre hydrodynamics \citep[e.g.][]{2008MNRAS.383..458C,2015MNRAS.450..277C,2018MNRAS.478.3544R,2025A&A...693A.180C} used 600~km~s$^{-1}$ for the stellar wind of IRS~16SW, based on its spectral features \citep{2007A&A...468..233M,2008MNRAS.383..458C}. 
    While relatively low, that velocity did not result in the formation of an unstable bow shock in the models. 
    IRS16SW however is a close binary, so its outflow velocity can be very different from that determined by its atmospheric properties. 
    High-resolution simulations of collisions of winds in static stellar pairs by \citet{2020MNRAS.493..447C} found that clumps formed by a symmetric binary tend to be ejected at $\approx$3/5 of the individual stellar wind velocity, so 360~km~s$^{-1}$ in this case.  While the orbit of the binary was not included in the model, we expect that its internal velocity (also around 360~km~s$^{-1}$) can be added to or subtracted from the outflowing material, resulting in a range of effective velocities for the outflow from 0 to $\approx$700~km~s$^{-1}$. 
    The low velocities could result in bow shocks as described above.
    
\begin{figure}[h]
        \centering
        \includegraphics[width=\linewidth]{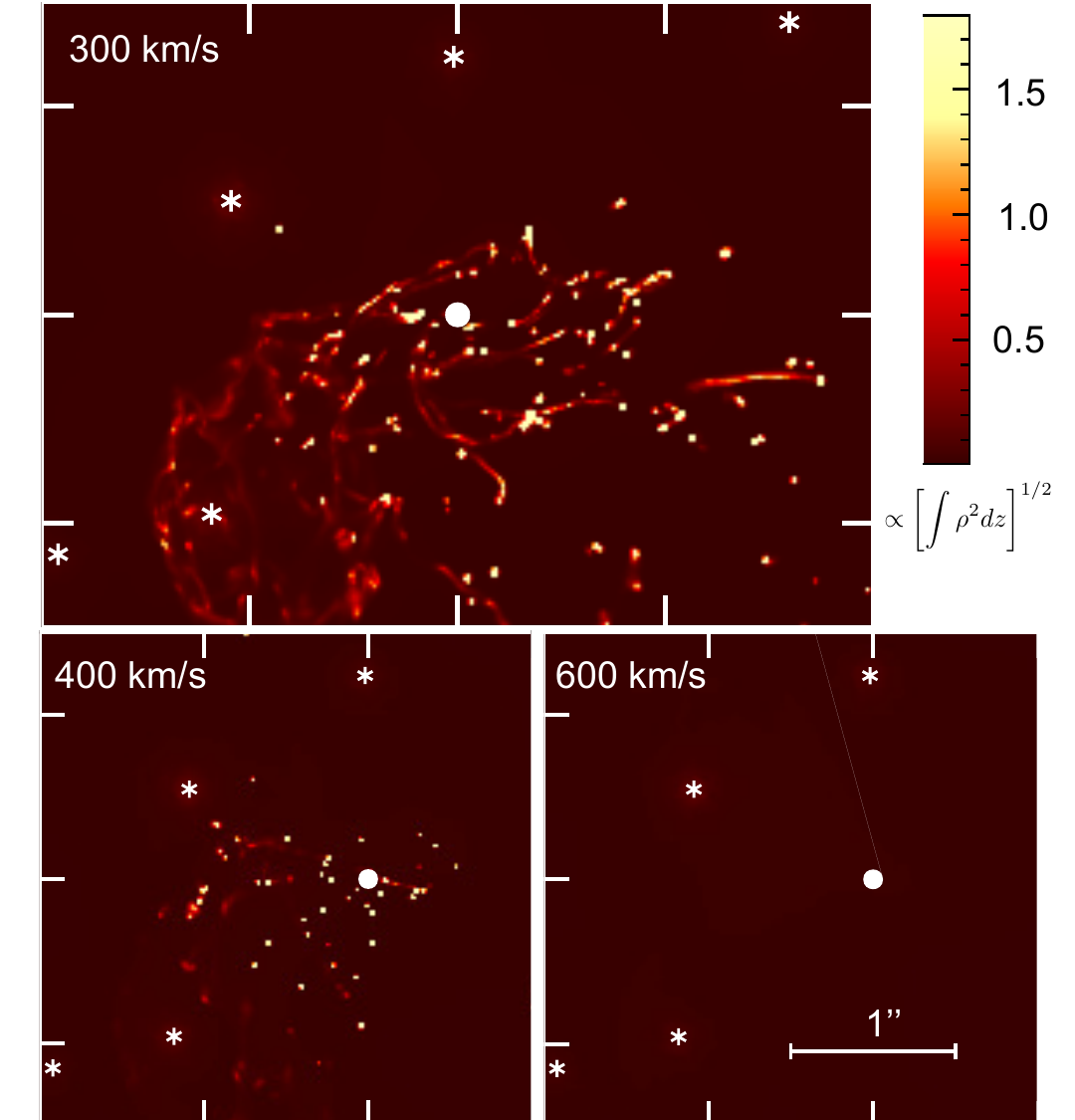}
        \caption{
        Snapshots of the hydrodynamic simulations of the Wolf-Rayet stars (white asterisks) feeding Sgr~A* (white disk) in the central parsec. 
        The maps show density square integrated along the the line of sight  in square root scale, i.e. $[\int\rho^2dz]^{1/2}$, represeting  the expected Brackett-$\gamma$ flux.
        Each panel represents a simulation run varying the wind speed of IRS~16SW: 300~km~s$^{-1}$ (top), 400~km~s$^{-1}$ (bottom left), and 600~km~s$^{-1}$ (bottom right). The production of clumps on the scale shown here is dominated by IRS16~SW.
        Snapshots created with Splash \citep{2007PASA...24..159P}.
        }
        \label{fig5}
    \end{figure}

    To test this idea, we ran a simulation of the Wolf-Rayet stellar winds feeding Sgr~A* using the setup of the control model (without feedback from Sgr~A*) of \citet{2015MNRAS.450..277C}. 
    The system of 30 mass-losing stars was placed on their observed orbits \citep{2008MNRAS.383..458C,2017ApJ...837...30G} with their constrained stellar wind properties \citep{2007A&A...468..233M,2008MNRAS.383..458C}, and evolved for 1100~yr but starting in the past in order to recover the state of the system at the present time.
  Different than in our previous work, we made use of the state-of-the-art smoothed-particle hydrodynamic (SPH) code Phantom \citep[Russell et al. in prep.]{2018PASA...35...31P} in order to avoid the formation of spurious clumps in the generic SPH technique \citep[e.g.][]{2013MNRAS.434.1849H}. 
    While ideally IRS~16SW should be modelled as a binary, its compactness makes that impossible with our current setup.
    We therefore explore the effect of its expected slower wind by modelling it as a single source with wind speeds of 300~km~s$^{-1}$, 400~km~s$^{-1}$, and 600~km~s$^{-1}$. 
    The latter being equivalent to previous works in the literature \citep[e.g.][]{2008MNRAS.383..458C,2015MNRAS.450..277C,2018MNRAS.478.3544R,2020ApJ...888L...2C}.  
    The simulations with 300~km~s$^{-1}$ and 400~km~s$^{-1}$ do result in the formation of a dense bow shock around this source, which gets disrupted into several clumps and filaments, some of which get in orbits towards Sgr~A* (see Figure~\ref{fig5}), staying roughly in the same orbital plane as the source.
    The clumps produced in the simulations also have masses of the same order of magnitude as the observed ones.  
    These results differ from the conclusions of \cite{2018MNRAS.478.3494C} because in the current models the clumps form in the bow shock in front of IRS~16SW's orbit, and not at the binary location itself.  The bow shock slows down the material and directs it preferentially in radial orbits towards Sgr~A*.
    So it seems possible that G1-2-3 originates from IRS16SW, although its exact origin would need to be determined from more detailed work.  

    Finally, it is worth noting that accreting a single G2-like clump (with a mass $\approx 1$ Earth mass) per decade suffices to explain the inferred accretion rate onto Sgr~A* of $\approx 10^{-7.6}~\text{M}_\odot~\text{yr}^{-1}$ estimated at the apocenter distance of G2 of  $\approx1000~\text{R}_\text{S}$ \citep{2014ARA&A..52..529Y,2019ApJ...871..126G}, i.e. the G1-2-3 streamer being decelerated by the drag force could be the main source of gas for Sgr~A* currently, if the gas  $\approx 50\,$yr after first pericenter settles into the accretion flow. This notion is in agreement with the generally accepted idea that Sgr~A* is mostly fed  directly by the hot plasma created by the stellar winds \citep{1999ApJ...517L.101Q, 2006MNRAS.366..358C}, but is even more specific. The process creating clumps happens preferentially when IRS~16SW approaches its pericentre, as it is currently happening, where the medium is denser, and it therefore not necessarily representative of a steady state. This could relate to the variable Sgr~A* emission on the time scales of decades to centuries as witnessed from X-ray reflection echoes \citep{2013A&A...558A..32C, 2018A&A...610A..34C}.

\section{Conclusions}
\label{sec:conclusions}

New near-infrared IFU data from the Galactic Center reveal a third object, looking alike and moving along (roughly) the same orbit as previously G1 and G2 did, although 18 years later. Given that by probabilistic arguments it is de facto excluded to find three stellar sources in such a configuration, the new data render a stellar model at least for these three sources very unlikely. A much better picture is that of a gas streamer G1-2-3, which appears to originate from a young, massive, binary star in the clockwise disk of stars - IRS16SW. The dominant differences between the orbits of the three gas clouds are the pericenter times and orientations of the (initial) orbital ellipses in their planes. These change systematically and synchronously with the motion of IRS16SW. We conclude that G1-2-3 most likely originates from the winds of that star. Updated hydrodynamic simulations show that taking into account lower wind velocities, as they occur around binaries, IRS16SW may well be able to produce clumps and filaments of gas that can reach Sgr~A*.

\begin{acknowledgements}
We are very grateful to our funding agency MPG, to ESO and the Paranal staff, and to the many scientific and technical staff members in our institutions, who helped to design, build, test, commission and operate ERIS.  Based on observations collected at the European Southern Observatory under the ESO programmes listed in appendix~\ref{sec:A}. The research of DC has been funded by the Alexander von Humboldt Foundation. JC acknowledges financial support from ANID -- FONDECYT Regular 1211429 and 1251444, and Millennium Science Initiative Program NCN$2023\_002$. The research of TP was supported by an advanced ERC grant 'multijets'.
\end{acknowledgements}

\bibliography{references}

\begin{appendix}

\section{Data}
  \label{sec:A}
  
\begin{table}[h]
 \caption[]{Integral-field spectroscopy observations used for G3}
 \label{tab:datalist}
 \begin{center}
 {\tiny
 \begin{tabular}{lcccc}
 Year&ESO prog. ID  &Instrument + Setup& total exposure\\
 \hline
2014 & 092.B-0398(ABD) & SINFONI, H+K-band & 1590 min\\
 & 093.B-0218(ABD) & SINFONI, H+K-band & \\
 &  093.B-0217(F)  & SINFONI, H+K-band & \\
2018 &299.B-5056(B) &SINFONI, H+K-band  & 1770 min\\
&598.B-0043(BDEFGHI) & SINFONI, H+K-band  & \\
& 0101.B-0195(BCDEFG) & SINFONI, H+K-band  & \\
2019&0103.B-0026(BDF) &SINFONI, H+K-band  &810 min\\
&5102.B-0086(Q) &SINFONI, H+K-band  &\\
&594.B-0498(Q) &SINFONI, H+K-band  &\\
2023&111.24H0.00(23)&ERIS, K-band  &770 min\\
2024& 112.25FZ.001 &ERIS, K-band  & 360 min\\
&  113.26B5.001 &ERIS, K-band  & \\
2025&114.2756.001&ERIS, K-band  & 340 min\\
\hline
 \end{tabular}
 }
\end{center}
\end{table}

\section{Orbital elements}
  \label{sec:C}
    We use the classical orbital elements for the non-Keplerian orbits in the sense of osculating elements. They are semi-major axis $a$, eccentricity $e$, inclination $i$,  position        angle         ascending   of the ascending    node $\Omega$, longitude       of the   pericenter $\omega$ and  epoch    of     pericenter      passage $t_P$. The mass $M$ of Sgr~A* and its distance $R_0$ are kept fixed. For the individual fits, we include the drag force parameter \citep{2019ApJ...871..126G} as a free parameter for G2, and use the resulting best-fit value as fixed quantity for the much less constrained G1 and G3 fits. The uncertainties are the formal fit uncertainties, rescaled such the reduced $\chi^2$ of each fit is 1. The orbit of G1 with our data is only poorly constrained, especially in $a$ and $e$, where the formal uncertainties don't properly represent the posterior space. Comparing with \citet{2017ApJ...847...80W} and/or including their data for increased phase coverage in our fits, leads to a semi-major axis value comparable with that of G2, and an eccentricity of around $e=0.98$. 
  \begin{table}[h]
 \caption[]{Orbit parameters for the individual orbit fits}
 \label{tab:oelist1}
 \begin{center}
 {\tiny
 \begin{tabular}{lccc}
 Parameter&G2&G1&G3\\
 \hline
$a$ [''] & $1.50 \pm 0.41$ & $0.23 \pm 0.06 $ &  $0.92 \pm 0.69$\\
$e$ & $0.989 \pm 0.003$ & $0.76 \pm 0.08$ &  $0.978 \pm 0029$\\
$i$ [$^\circ$]& $122.0\pm 0.8$ & $112.8 \pm 3.9$ &  $126.1 \pm 8.7$\\
$\Omega$ [$^\circ$]& $70.0\pm 3.3$ & $59.0 \pm 8.1$ &  $49.7 \pm 49.3$\\
$\omega$ [$^\circ$]& $94.6\pm 2.3$ & $97.41 \pm 8.8$ &  $91.5 \pm 33.1$\\
$t_P$ &$2014.18\pm0.12$ &$2002.8\pm 0.7$ &$2033.6 \pm 2.5$ \\
\hline
$\chi^2_\mathrm{red}$ & 2.31 & 2.76 & 1.89 \\
 \hline
 \end{tabular}
 }
\end{center}
\end{table}

\begin{table}[H]
 \caption[]{Orbit parameters for the common orbit fit}
 \label{tab:oelist2}
 \begin{center}
 {\tiny
 \begin{tabular}{lcc}
 Parameter&Value&formal fit error\\
 \hline
$a$ [''] & 0.9540 & 0.0810 \\
$e$ [''] & 0.98111 & 0.0019 \\
$i$ [$^\circ$] &120.14&0.35 \\
 $\Omega$   [$^\circ$] &77.50&0.64 \\
  $\omega$ G2  [$^\circ$] &99.52&0.56 \\
   $t_P$ G2 [yr] & 2013.945& 0.042\\
$\omega$   G1  [$^\circ$] &91.57&0.91 \\
   $t_P$ G1 [yr] & 2003.123& 0.075\\
 $\omega$  G3  [$^\circ$] &112.44&0.88 \\
   $t_P$ G3 [yr] & 2031.513& 0.206\\
 \hline
 \end{tabular}
 }
\end{center}
\end{table}

\section{Additional illustration}
\begin{figure}[h]
\centering
\includegraphics[width=9cm]{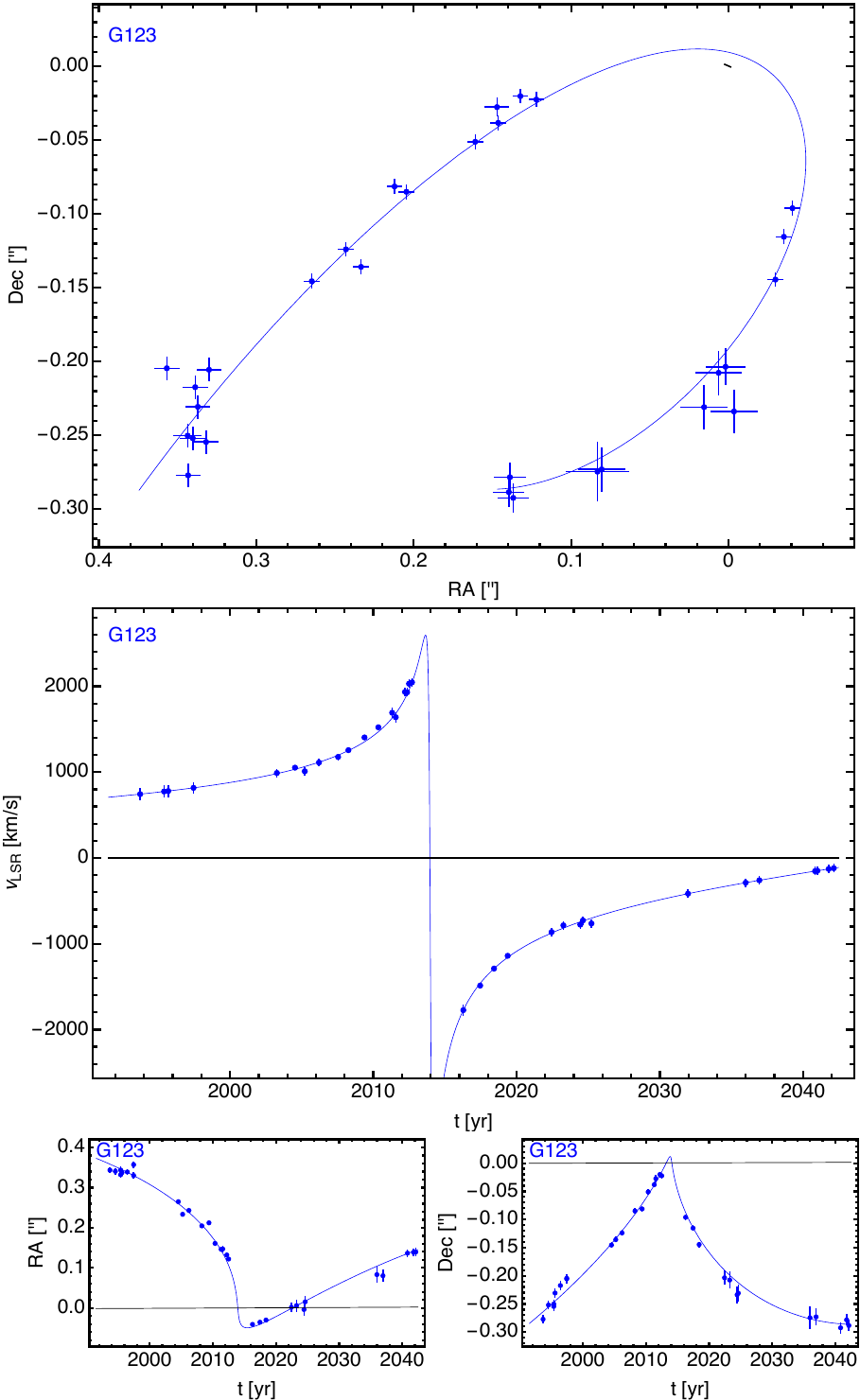}
\caption{Alternative representation of the combined orbit fit for G1, G2 and G3. The data from G1 and G3 have been corrected for the differences in the respective orbits to the G2 orbit, and plotted on top of the G2 orbit and data.}
\label{fig10}
\end{figure}

\end{appendix}

\end{document}